\documentclass[ conference,final]{IEEEtran}

\IEEEoverridecommandlockouts  %thanks command only

    \usepackage{amsmath} 
    \usepackage[utf8]{inputenc} 
    \usepackage[T1]{fontenc}
\usepackage[colorlinks=true,
        linkcolor=red,
        urlcolor=blue,
        citecolor=blue]{hyperref}
\usepackage{cite}
\usepackage{graphicx, verbatim}
\usepackage{ccicons}

\usepackage{booktabs}
\usepackage{textcomp}
\usepackage{array}

\title{\LARGE \bf
Cluster-based Approach to Improve Affect Recognition from Passively Sensed Data
}

\author{Mawulolo K. Ameko$^{1}$, Lihua Cai$^{1}$, Mehdi Boukhechba$^{1}$, Alexander Daros$^{2}$,\\ Philip I. Chow$^{3}$, Bethany A. Teachman$^{2}$, Matthew S. Gerber$^{1}$, Laura E. Barnes$^{1}$
\thanks{$^{1}$Department of Systems and Information Engineering,
        University of Virginia, Charlottesville, VA 22904, USA
        }%
\thanks{$^{2}$Department of Psychology, University of Virginia, Charlottesville, VA 22904, USA
}%
\thanks{$^{3}$Department of Psychiatry and Neurobehavioral Sciences, University of Virginia, Charlottesville, VA 22904, USA
}
}

\begin{document}

\maketitle
\thispagestyle{empty}
\pagestyle{empty}

\begin{abstract}
Negative affect is a proxy for mental health in adults. By being able to predict participants' negative affect states unobtrusively, researchers and clinicians will be better positioned to deliver targeted, just-in-time mental health interventions via mobile applications. This work attempts to personalize the passive recognition of negative affect states via group-based modeling of user behavior patterns captured from mobility, communication, and activity patterns. Results show that group models outperform generalized models in a dataset based on two weeks of users' daily lives.
\end{abstract}

%\cite{watson2014personality}
\section{Introduction}
The extent to which individuals experience positive and negative affect on a daily basis is associated with mental health outcomes~\cite{clark1994temperament}. Higher levels of negative affect are associated with increased vulnerability to many mental disorders, including depression and anxiety disorders, two of the most common types of mental disorders in U.S. adults~\cite{kessler2005lifetime}. Mental health research typically relies on self-report questionnaires that assess negative affect at a moment in time. Repeated administration of these measures, such as in an ecological momentary assessment (EMA) framework, is resource intensive and susceptible to retrospective bias when participants are asked to recall their mood over a previous duration \cite{gentzler2006adult}. Ideally, negative affect would be recognized without asking participants, thereby reducing burden, improving compliance among participants, and allowing for continuous modeling of affect change. To aid recognition of negative affect, unobtrusive mobile sensing of location, texts and calls, and activity levels could also be used to enrich the information provided by participants' responses to questionnaires assessing negative affect and measures of mental health (e.g., social anxiety, depression). 

Current affect recognition approaches are based primarily on generalized or individualized approaches \cite{yonekura2015mood}. In generalized approaches, the recognition model learns global patterns that the majority of participants followed during the experiment. These patterns are then used for prediction. Since user behaviors vary substantially, generalized models may fail to predict variations in affect for an individual person. In contrast, individualized models are designed to learn participants' patterns on a case-by-case basis, thus they are expected to be more accurate. However, individualized models require a certain number of observations for each individual to obtain robust prediction performance. In short-term studies involving human subjects (e.g., two weeks), individual models may fail to adequately capture individual affective patterns because of a small pool of observations~\cite{likamwa2013moodscope}. 

In our work, we propose a new group-based approach that integrates generalized and personalized models. We first propose a method for clustering multimodal behavioral profiles that groups participants based on their mental states, activity levels, communications, and mobility patterns. We then apply several prediction algorithms to investigate whether group models using multimodal user profiles outperform the generalized or population-based model.

%some of our own works that may be relevant for citation.
%huang2016assessing
%chow2017using
%xiong2016sensus -- data collection platform

%what is affect recognition and its importance in mental health management, prevention, and treatment.

%what have been done in a high level summary
%what are still remaining as challenges to be solved

%the underpinning/connection between smartphone passive sensing data and users' personal traits

%what we are contributing in this work

\section{Related Work}
Smartphone usage can be used as an indirect marker of mood. Passively sensed location information has been used to predict depressive symptoms \cite{Saeb_2015}. Individuals with higher social anxiety levels were more likely to report negative affect during the day, which in turn was predictive of spending more time at home at subsequent measurements~\cite{chow2017using}. Self-reported stress and mental health indices were also successfully predicted in a 10-week long study design in college students with both passively and actively sensed data~\cite{Wang_2017}. 

Prediction of affect from mobile sensing appears to be more difficult to replicate. In a feasibility study, LiKamWa et al.~\cite{likamwa2013moodscope} explored a  personalized feature selection approach to predict changes in mood from unobtrusively sensed indices of social activity (e.g., calls/texts, emails), physical activity~(e.g., GPS), and general mobile phone use (e.g., application use, web browsing). The study relied on two months of data collected from 32 participants. Results indicated high levels of accuracy in predicting mood using personalized models. The personalized modeling also produced better accuracy compared to a generalized model using data from all users. 

A follow-up study in which a personalized feature selection approach was used to predict affect ratings from~$27$ participants over~$42$ days found no clear benefits of using this approach~\cite{asselbergs2016mobile}. However, these studies did differ in length, participant variability (e.g., depressive symptoms), and unobtrusive features assessed. It remains possible that personalized feature selection requires an intensive level of data collection that participants may perceive as burdensome. Given these findings, we use an intermediate approach between generalized and personalized models to recognize affect in a given situation.
%brief summary about what has been done by researchers in current literature in affect recognition

%for the problem that we are attempting address, what has been done

%compare and contrast throughout this section between our approach and others' approach

\section{Study Design}
\begin{figure}[htbp!]
\begin{center}
  \includegraphics[width=.9\columnwidth, height = 4.1cm]{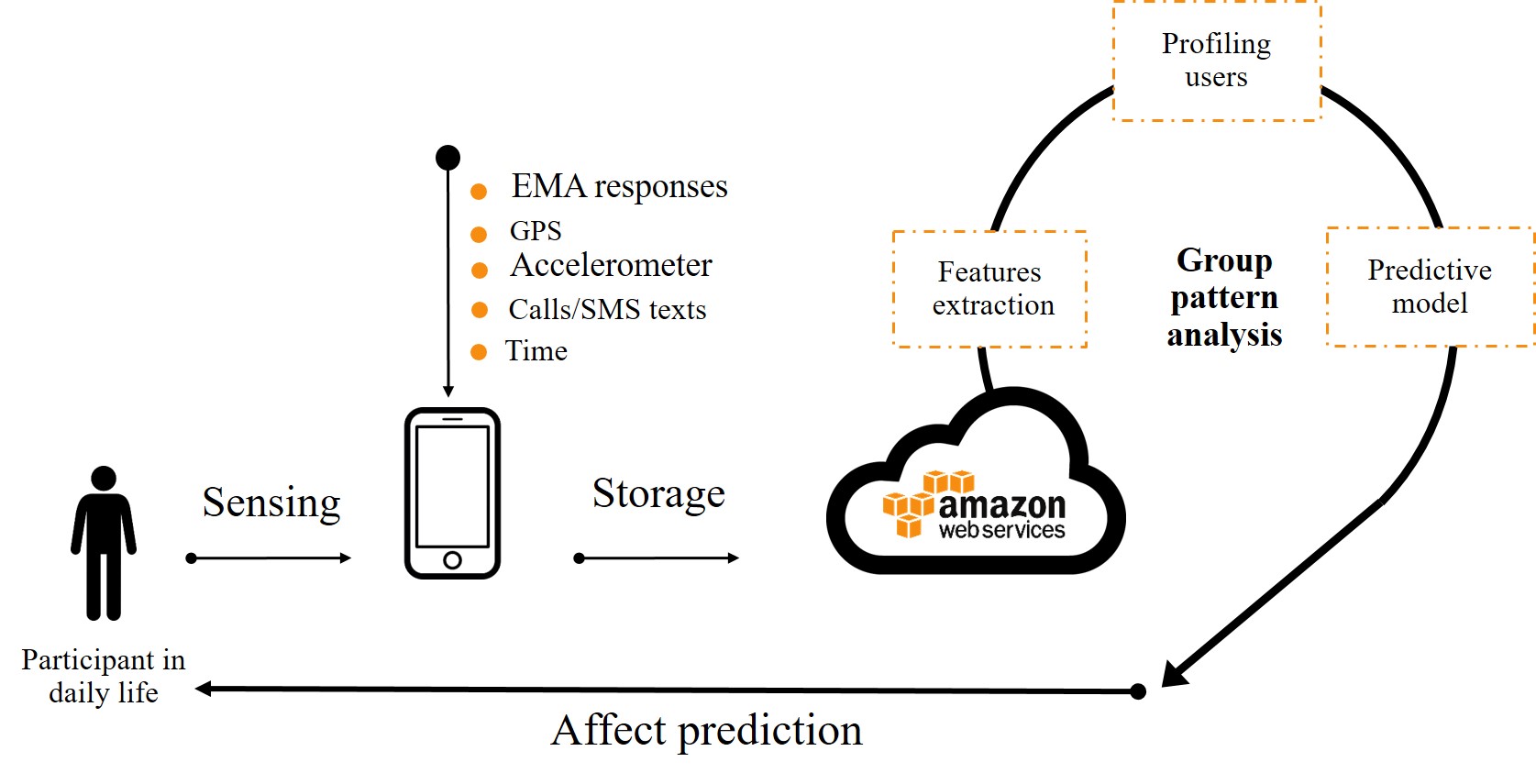}
  \caption{Passive and affect data collection using smartphones.}~\label{fig:overall_design}
  \end{center}
  \vspace{-15pt}
\end{figure}

Sixty-five undergraduate students were recruited for a two-week study period to understand dynamics of emotional, cognitive, and interpersonal processes associated with depression and social anxiety. University students provide a relatively homogeneous sample in terms of life phase and common psychological stressors, thereby mitigating the impact of a wide variety of potential nuance factors. Pre-study surveys were given to the students at enrollment, and one of these surveys measured students' social anxiety (SIAS)~\cite{mattick1998development}. The study contained an ecological momentary assessment (EMA) phase that requests self-report data on psychological affect throughout the day. A customized mobile app (Sensus) \cite{Xiong_2016} was installed on participants' personal Android smartphones and was programmed to deliver 6 EMAs throughout the day~(each survey contained 12 questions), randomly scheduled in each~2-hour block from~9~a.m. to~9~p.m. (e.g., once between 9-11 a.m., once between 11 a.m.-1 p.m., etc.). Sensus was also configured to deliver an end-of-day survey at 10 p.m. each day. Prompts concerning affect first asked participants to rate how positive they were feeling from~1~(not at all) to~100~(very positive). The second question asked participants to rate how negative they were feeling from~1~(not at all) to~100~(very positive). In addition to these active assessments, Sensus also passively collected GPS coordinates every 150 seconds and accelerometer data at~1~Hz, in addition to call and text logs.  All data were transmitted wirelessly to a secure Amazon Web Services server, where data were stored for further analysis~(see Figure~\ref{fig:overall_design}).

\section{Experiments}

\subsection{Data Preprocessing}
We first processed participants' raw GPS data into semantic locations (e.g., leisure, education, and home) by combining a spatiotemporal clustering algorithm ~\cite{kang_extracting_2005} and OpenStreetMap (OSM) geodatabase ~\cite{shekhar_openstreetmap_2015}. Our label taxonomy includes the following types: Education (e.g., university and libraries), Leisure (e.g., restaurants and cinemas), Out of town, In transition (e.g., going from one place to another), Home, and Other houses. Our algorithm has been trained to recognize Home as the place having a house OSM-tag (e.g., apartment, dormitory, house, etc. See ~\cite{shekhar_openstreetmap_2015} for more details about OSM tags) where a subject stayed the most between 10 p.m. and 9 a.m. 

For accelerometer data, we used statistical measures (mean, minimum, maximum, standard deviation, median and variance) on the 1-minute sliding window to extract several features of phones' motion around affect assessment moments. These features aim to represent the physical activity levels of the participants, and we used them to predict momentary negative affect. Note that our accelerometer features are extracted from the magnitude of acceleration $\sqrt{\frac{x^2+y^2+z^2}{3}}$
to make them orientation free, since the phones were used in participants' natural environments.

Individuals' affect may be associated with the degree to which they interact with others. Thus, we included communication events in our models. For each EMA we collected the number of text messages and phone calls as long as their duration overlapped with epochs prior to the EMA prompt. Here we chose 1 hour prior to the EMA prompt as the time window to record the number of text messages and phone calls.

\subsection{Profiling Users}

After preprocessing the data, we clustered the participants based on their behavioral profiles. There are different ways to cluster participants. For instance, a clustering strategy can be based on time spent at home to cluster people having depressive symptoms, drawing on the hypothesized correlation between home staying and affect fluctuation patterns. The following four passively sensed profile features were used to drive the clustering process.

%Participants' profiles based on their locations, activity levels (e.g. acceleration from accelerometer data), and communications (e.g., number of text messages and calls) in different epochs prior to the EMA prompt.

\subsubsection{Location}
For location data, we considered five common point-of-interest classes consisting of \{`out of town', `education', `friends' houses', `home', `leisure'\}. Then we calculated the proportion of time spent in each of these locations over the study period for each participant. 

\subsubsection{Activity}
From the accelerometer data, we chose thresholds of 0.2 and 0.3 between the minimum and maximum to define three levels of activity (e.g., \{Low, Medium, High\} in acceleration). We chose these cutoffs based on the observed distribution of the acceleration values. Then for each participant, we calculated the proportion of time being in these activity levels (e.g., proportion of time being in the high level).

\subsubsection{Short-Message Service (SMS)}
From the SMS data, we aggregated the number of text messages sent and received within each 1-hour window during the study period. From this, we defined 5 text messaging levels based on text message frequencies (e.g., `VeryLow',`Low',`Medium',`High',`VeryHigh')  with intermediary cutoffs at 1, 10, 20 and 30 messages per hour based on their observed distribution.

\subsubsection{Phone Calls}
Similarly, we computed the proportion of calls occurring at each level of call activity defined as `Low',`Medium',`High',`VeryHigh' using thresholds of 1, 3 and 6 calls per 2-hour window. We used a 2-hour window to accommodate the lower hourly frequency of phone calls compared with text messages.

%\subsubsection{SIAS score}

Formally, for the design matrix $X \in \mathbf{R}^{N \times d}$ with $X=\{\mathbf{x_i}\}_{i}^{N}$, the feature vector for each participant is $\mathbf{x_i} = \{ \underbrace{x_{i1},x_{i2}, \cdots, x_{ip_1}}_\textrm{M1}, \underbrace{x_{i1},x_{i2}, \cdots, x_{ip_2}}_\textrm{M2}, \cdots, \underbrace{x_{i1},x_{i2}, \cdots, x_{ip_n}}_\textrm{Mn}\}$. Note that $\textrm{M}_{i}$ ($ i \in [1,n]$) represents the $i$th modality and $p_{i}$ the number of levels in the $i$th modality.% In our case, $n=4$, $N=64$, and $d=?$.

With the above, we determine different clusters based on various combinations of these four passively sensed modalities in addition to SIAS using the G-means~(Gaussian Means)~\cite{hamerly2004learning} algorithm. The G-means algorithm is an extension of K-means where number of clusters is automatically determined by iteratively selecting $k$ such that the data assigned to each cluster follows a Gaussian distribution. %We believe that there does not exist a predetermined number of clusters of participants in our cohort as is usually the case for most clustering methods.

\subsection{Predictive Models}
We used 4 algorithms to test the predictability of negative affect: Gaussian process, SVM, linear lasso, and random forest. Each of these models has merit with respect to the issues that may ensue from constraints of data availability for model training, which is the case in this study. Although random forest, SVM, and Lasso regression are well-studied, Gaussian processes have demonstrated promising performance in e-health applications \cite{clifton2013gaussian} mostly because they enable experts to encode their beliefs about smoothness or periodicity using covariance functions. In addition, the complexity of the model is inherently regulated (see chapter 5 of \cite{rasmussen2006gaussian}) and provides uncertainty over predicted values. In our case, we used the squared-exponential covariance function \cite{rasmussen2006gaussian}:
\begin{equation}
K(x,x') = \theta^2_{s} exp\left \{\frac{-||x - x'||^2}{2\theta^2_{\ell}} \right \}
\end{equation}
where $\theta_t = \{ \theta_{s}, \theta_{\ell} \}$, with $\theta_{s}$ and $\theta_{\ell}$ being the hyperparameters of the covariance function regulating the y-scale and x-scale, respectively.

\section{Results}

\begin{figure}[htbp!]
\begin{center}
  \includegraphics[width=\columnwidth, height = 4.1cm]{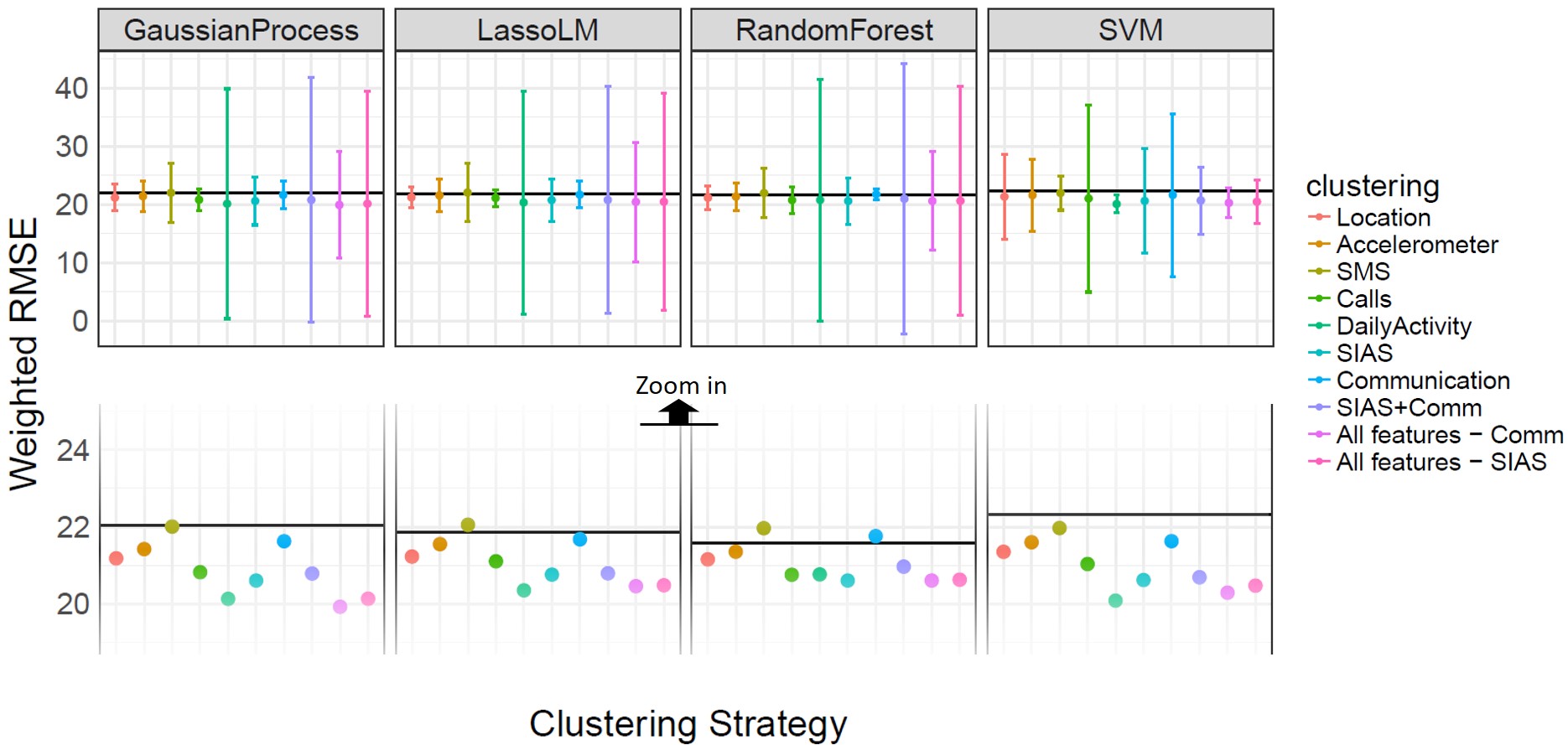}
  \caption{The performance of each grouping strategy compared with the generalized model’s performance (black horizontal line). The y-axis is the weighted root mean square error (WRMSE). The error bars represent 2 standard deviations of each grouping strategy.}~\label{fig:overall_performance}
  \end{center}
  \vspace{-15pt}
\end{figure}

%\subsection{Multimodal Clusters}
Figure \ref{fig:overall_performance} presents the performance of various clustering strategies compared with generalized models using the predictive algorithms presented earlier. Before analyzing performance, we will present a brief interpretation of each grouping strategy. Using data from SMS, four groups were discovered as presented in Table \ref{clustering_component1}. The group labeled \textit{freq} are most actively engaged with text messaging on their phones, while \textit{reg1} and \textit{reg2} fall in the middle with \textit{reg2} being more frequent than \textit{reg1}. The most inactive group is labeled by \textit{infreq}. In the profiles learned using the phone call logs, two groups were discovered: an active group and an inactive group in terms of their phone call level distributions. Notice that for the majority of time prior to EMAs, phone calls were rarely made by our study participants, and thus we see high percentages in the `low' level. Using acceleration as a proxy to characterize participants' activity level, we found two: one \textit{active} group and one \textit{inactive} group. Again notice that the differences in the acceleration level distribution between the two learned groups are minor and only relative between them. With respect to locations, in the first group, the participants split most of their time between school and home; in the second group, the participants spent over 80\% of their time at school at the expense of other places; and in the third group, the participants spent the majority of their time away from home (e.g., traveling out of town, visiting friends, and at leisure place of interests).

We also used cutoffs of 34 and 43 in SIAS scores to divide participants into low, medium, and high social anxiety groups~\cite{heimberg1992assessment}. In total, we experimented with 10 grouping approaches based on location, activity level, communications~(SMS and phone calls), and SIAS scores as shown in Figure~\ref{fig:overall_performance}. Specifically, \textit{DailyActivity} applies a combination of location, activity level, communications profiles; communication is based on the combination of phone calls and SMS~(re-grouped into active and inactive) producing three groups~(active in both SMS and calls, only active in either SMS or calls, inactive in both SMS and calls). 

\begin{table}\footnotesize
\begin{center}
\caption{Clustering based on communication, location, and acceleration data using G-means clustering algorithm.}
\label{clustering_component1}
\begin{tabular}{ p{0.05\columnwidth} p{0.05\columnwidth} p{0.08\columnwidth} p{0.05\columnwidth} p{0.05\columnwidth} p{0.05\columnwidth}  p{0.05\columnwidth}  p{0.05\columnwidth}  p{0.05\columnwidth} } 
 \toprule
 & & & & \multicolumn{5}{c}{Group Profile (\%)} \\
 \cmidrule{5-9}
 & Gp & Label & \#Part & Low+ &  Low & Med & High & High+  \\ [0.5ex] \midrule
SMS & 1 & reg1 & 22 & 80.5 & 16.8 & 2.0 & 0.5 & 0.2 \\
 & 2 & reg2 & 12 & 68.6 & 25.9 & 4.3 & 0.7 & 0.4 \\
 & 3 & infreq & 9 & 93.7 & 5.9 & 0.3 & 0.1 & 0.0\\
 & 4 & freq & 19 & 49.1 & 36.1 & 8.6 & 3.5 & 2.7\\
\midrule
Call & 1 & inactive & 54 & & 89.5 & 9.1 & 1.3 & 0.1 \\
 & 2 & active & 8 & & 65.7 & 29.2 & 4.4 & 0.7 \\
 \midrule
Acc & 0 & active & 25 & & 83.1 & 4.6 & 12.3 & \\
 & 1 & inactive & 37 & & 91.2 & 2.7 & 6.1 &  \\
%\bottomrule
 \toprule
%  & & & \multicolumn{5}{c}{Group Profile (\%)} \\
% \cmidrule{4-9}
 &  &  &  & Out &  Edu & Friend & Home & Leisure \\ [0.5ex] 
 \midrule
Loc & 1 & school-home & 34 & 2.0 & 49.2 & 4.4 & 38.7 & 5.8 \\
 & 2 & school & 18 & 3.0 & 83.0 & 2.7 & 4.9 & 6.5 \\
 & 3 & out & 10 & 20.9 & 43.7 & 8.3 & 9.3 & 17.8\\
 \bottomrule
\end{tabular}
\end{center}
\end{table}

\begin{comment}
\begin{table}\footnotesize
\begin{center}
\caption{Clustering based on place visit data GMeans clustering algorithm.}
\label{clustering_component2}
\begin{tabular}{ p{0.03\columnwidth} p{0.17\columnwidth} p{0.05\columnwidth} p{0.05\columnwidth} p{0.05\columnwidth}  p{0.06\columnwidth}  p{0.05\columnwidth}  p{0.1\columnwidth} } 
 \toprule
  & & & \multicolumn{5}{c}{Group Profile (\%)} \\
 \cmidrule{4-8}
 Gp & Label & \#Part & Out &  Edu & Friend & Home & Leisure \\ [0.5ex] 
 \midrule
 1 & school-home & 34 & 2.0 & 49.2 & 4.4 & 38.7 & 5.8 \\
 2 & school & 18 & 3.0 & 83.0 & 2.7 & 4.9 & 6.5 \\
 3 & out & 10 & 20.9 & 43.7 & 8.3 & 9.3 & 17.8\\
 \bottomrule
\end{tabular}
\end{center}
\end{table}
\end{comment}

%interpret these results
%if space allowed, create two dimensional plots on each clustering component using the first two PCA components. Turn these plots into one figure.

%get better names for these groups in each component.

%\subsection{Generalized vs. Group Models}

From Figure~\ref{fig:overall_performance}, using most of the grouping strategies, we were able to obtain better overall performance in lower weighted RMSE in our group models when compared to the generalized model. Specifically, our generalized models using four different algorithms achieved a RMSE of~$21.58$~(random forest), $22.05$~(Gaussian processes),~$21.87$~(linear lasso), and~$22.31$ (SVM), respectively. %On average, we obtained a reduction of xxxx in RMSE as in WRMSE of different grouping strategies across all grouping strategies and all four algorithms.
For each grouping strategy on Gaussian processes model, we were able to obtain average reductions of RMSE~$0.8722$~(Location),~$0.6310$~(activity level),~$0.045$~(SMS),~$1.2330$~(calls),~$1.4505$ (SIAS), $1.9268$~(DailyActivity),~$0.4264$ (communication),~$1.2675$~(SIAS+communication), $2.1326$ (All features - communication), $1.9231$ (All features - SIAS), respectively.

%What is the paln here?
%Maybe an opportunity to make use of the plots provided by Lihua?

%\subsection{Behaviors on Group Model Performance}

Note from Figure~\ref{fig:overall_performance} that the DailyActivity grouping strategy consistently performed better than most other grouping strategies, and this strategy is also closest to the individual model approach (65 individual models for 65 participants) because it resulted in the most (25) subgroups among all these strategies, thus we used it to further investigate whether there are any specific patterns with respect to sample size to guide future design of group-level modeling approaches.

\begin{figure}[htbp!]
\begin{center}
  \includegraphics[width=\columnwidth, height = 4.1cm]{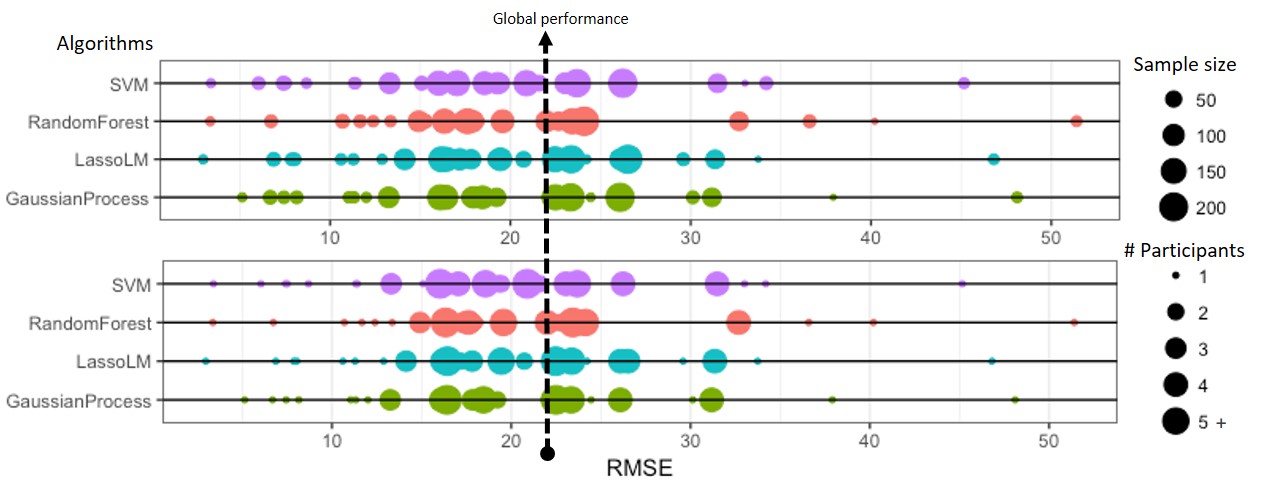}
  \caption{The impact of sample size on the performance of groups formed by DailyActivity strategy.}~\label{fig:samplesize_perf}
  \end{center}
  \vspace{-10pt}
\end{figure}

From Figure~\ref{fig:samplesize_perf}, we can see that there is a nonlinear relationship between sample size of groups and their performances. Groups with small sample size tend to perform either extremely poorly or extremely well. This signals potential weak generalizability of profiling strategies that forms many small groups. So the ideal situation will be to form groups with profiling strategies that evenly distribute the samples across different subgroups.

%\subsubsection{Activity Level, Location, and Communication} We conducted analysis of variance (ANOVA) on the RMSE of each subgroup on each of these four factors (activity level, location, and communication groups shown in Table~\ref{clustering_component1} using the DailyActivity profiling strategy, and found no significant difference among the subgroups. Although no significant differences on RMSE level of different algorithms, further investigations using larger dataset is required to validate this conclusion.% Note that our groups within each clustering strategy is automatically determined by the GMeans clustering algorithm, different studies may learn different groups.

%\subsubsection{SIAS} For SIAS profiling, although not included in the DailyActivity profiling strategy (lest we explode the number of subgroups), has wide applicability because of the popularity of pre-study surveys being administered in EMA studies. %From the performance in RMSE across all four algorithms, we observed that RMSE follows the high, low, and medium social anxiety order. This means that negative affect in higher and lower social anxiety students is more predictive than those with moderate level of social anxiety. 

%what criteria are important to gauge the performance of a group level models?

%in general, how to form clusters? what insights can we get based on our data/experiments/analysis?
%sample size on rmse/performance
%acc
%loc
%communication - sms, calls

%why some clustering approach fails? why most of them work?

%when forming the clusters, it shouldn't be more clusters than the number of participants ...

\section{Conclusion}
The focus of the present investigation was to provide a framework for accurately predicting negative affect from passively sensed data concurrent with individuals' affect ratings. Given that two weeks may be too short for algorithms to learn personalized models, we developed a method for predicting negative affect using a group-level approach. We first clustered participants using multimodal behavioral profiling, then we predicted negative affect from passively sensed data. The results indicate that profiling users based on their behavior improves the performance of the predictive model compared to generalized models. Future work will study the predictability levels among the different groups using validated questionnaire measures of personality and depression. The present study contributes to a body of research that aims to use passively sensed data to recognize user affect and launch interventions when and where they are most needed.

\section*{ACKNOWLEDGMENT}

This research was supported by the Hobby Postdoctoral and Predoctoral Fellowships in Computational Science, and NIMH R34MH106770 and NIMH R01MH113752 grants. 

\bibliographystyle{IEEEtran.bst}
\bibliography{literature}

\end{document}